\documentclass[twocolumn,aps,superscriptaddress,pre]{revtex4}

\usepackage{amsmath,amssymb,graphicx}
\usepackage{subfigure}
\usepackage{algorithmic}
\usepackage{enumerate}
\usepackage{times}
\usepackage{color}
\usepackage{soul}
\definecolor{yblue}{rgb}{0.06, 0.3, 0.57}
\usepackage[pdftex]{hyperref}
\hypersetup{colorlinks=true,linkcolor=blue,citecolor=blue,urlcolor=blue}

\DeclareMathOperator{\sech}{sech}

\begin{document}

\title{Controlled engineering of a vortex-bright soliton dynamics using a constant driving force}

\author{Wenlong Wang}
\email{wenlongcmp@scu.edu.cn}
\affiliation{College of Physics, Sichuan University, Chengdu 610065, China}

\begin{abstract}
A vortex-bright soliton can precess around a fix point. Here, we find numerically that the fixed point and the associated precessional orbits can be shifted by applying a constant driving force on the bright component, the displacement is proportional to the force with a minus sign. This robust dynamics is then discussed theoretically by treating the vortex-bright soliton as an effective point particle, explaining the observed dynamics and predicting new ones that are subsequently confirmed. By appropriately tuning the force, the vortex-bright soliton can be guided following an arbitrary trajectory, including that it can be pinned and released at will. This finding opens a highly flexible and controllable approach of engineering the dynamics of vortical structures in Bose-Einstein condensates.
\end{abstract}

\maketitle

\section{Introduction}

Bose-Einstein condensates (BECs) have provided a highly controllable setting for the study of solitary waves, particularly dark solitons \cite{Dimitri:DS}, vortices \cite{Alexander2001,fetter2}, and their higher-dimensional generalizations, e.g., vortex rings and knots \cite{becbook1,becbook2,PhysRevE.85.036306}. The fascinating vortex state, which has a localized density defect but a long-range phase winding, has many peculiar properties and has attracted considerable attention \cite{Hall:VX2,PK:VXreview}. Vortices are also widely studied in superfluids \cite{Nicolay:SF}, superconductors, and nonlinear optics \cite{DSoptics}, from the perspectives of topological quantum computing \cite{VXQC}, and even gravity \cite{Jeff:HR}. Vortex generation and dynamics play an important role in the dissipation of superfluid transport and turbulence \cite{Dissipation}. The proliferation of vortex-anti-vortex pairs is central to the superfluid-normal fluid transition \cite{Nicolay:SF}. Vortices can be controllably generated in BECs using the phase imprinting technique \cite{VB,DBS2,Panos:BEC3C}, and other methods are also available such as rotating or stirring a condensate \cite{Madison:VX}, through matter-wave interference \cite{MWI1,MWI2}, and through the dynamical instabilities of dark soliton structures, e.g., the ring dark soliton \cite{Wang:AI}. Importantly, vortices in BECs can be observed in real time \cite{Hall:Realtime}.

The vortex structure has a two-component counterpart, the vortex-bright (VB) soliton, in the context of multicomponent BECs \cite{VB,MWIVB}. Interestingly, the vortex in BECs was firstly generated in the form of a VB using phase imprinting \cite{VB}. We focus on this structure in this work. Here, a bright soliton of one component is trapped by the vortex core of the other component \cite{PK:VB} despite that a bright soliton cannot exist on its own in a one-component repulsive condensate. The idea that a density dip of one component can trap another component transcends dimensionality. Particularly, the 1d sibling is the extensively studied dark-bright (DB) soliton \cite{DBS1,DBS2,revip}, where a bright soliton is instead trapped by a dark soliton. Vector solitary waves can be implemented using a spatially dependent spin interconversion and phase imprinting in pseudo-spinor and even spinor condensates in, e.g., $^{87}$Rb \cite{DynamicsF2,Chapman:spinor,Chapman:np,Panos:BEC3C} or $^{23}$Na condensates \cite{Ketterle:sd,ketterle:spinor} in far detuned optical dipole traps.

It was recently demonstrated that the trapped bright soliton provides an excellent opportunity for engineering the dynamics of a vector soliton by driving the bright soliton using a constant driving force \cite{Lichen:DB,Wang:MDDD}. Naturally, engineering soliton dynamics is of enormous interest, and previous works have already paid much attention to guiding or driving solitons with time-varying potentials, particularly optical lattices and impurity potentials \cite{BSIM,DSOL,VXIM}. Here, our approach differs in that we only tune the driving force to piecewise design the VB trajectory, i.e., there is no adiabatic dragging of a soliton in the process. The driving-induced AC oscillation of the 1d DB soliton was examined in detail in \cite{Lichen:DB}, finding that the DB soliton initially propagates against the driving potential and then oscillates back, and so on. This peculiar dynamics was understood from the dark soliton effective negative mass and the Hamiltonian nature of the system. However, the dynamics has not been explored in the 2d VB setting.

The main purpose of this work is to investigate the VB driving-induced dynamics numerically, describe the dynamics theoretically, and finally explore the VB dynamics engineering. First, we find interestingly that the VB dynamics differs significantly from that of the DB counterpart. The VB executes uniform circular motion with very different scaling properties and mechanisms, which is traced to the long-range nature of its phase winding, despite that it has a short-range density defect. In addition, the fixed point of the VB is not destroyed but rather shifted by the force, enabling an excellent opportunity of VB dynamics engineering. As demonstrated in Sec.~\ref{results}, we can guide the VB soliton following essentially an arbitrary trajectory, including that it can be pinned at a desired position and released at will. Similarities and differences of the driving-induced VB and DB dynamics are also discussed in this work.

This work is organized as follows. In Sec.~\ref{setup}, we introduce the model and the numerical setup. Next, we present our numerical results and theoretical analysis in Sec.~\ref{results}. Finally, our conclusions and future considerations are given in Sec.~\ref{conclusion}.

\section{Computational setup}
\label{setup}

Our starting point is the well-known coupled dimensionless Gross-Pitaevskii equation \cite{Panos:book}:
\begin{eqnarray}
-\frac{1}{2} \nabla^2 \psi_j + V_j \psi_j + \left(\sum_{k=1}^2 g_{jk}| \psi_k |^2\right) \psi_j = i \frac{\partial \psi_j}{\partial t},
\label{GPE}
\end{eqnarray}
where $\psi_j(x,y,t), j=1, 2$ are two complex scalar macroscopic wavefunctions. Motivated by the earlier DB work \cite{Lichen:DB}, we use the interaction coefficients $g_{11}=3$, $g_{12}=g_{21}=2$, and $g_{22}=1$. In this setting, $\psi_1$ is the vortex state and $\psi_2$ is the bright soliton state. The advantage of these coefficients is that there exists VB structures with approximately a uniform total density $|\psi_1|^2+|\psi_2|^2$. This significantly simplifies the theoretical analysis as the interaction energy is conserved to an excellent approximation. The dynamics, however, is not limited to this particular choice of interactions \cite{Wang:MDDD}.

We work with a finite-size homogeneous system, the condensates are confined in the following effectively hard-wall potential \cite{BoxP} $V(r)$:
\begin{align}
\label{Vr}
V(r) &= U_w \big( \tanh(r-R_w)+1 \big)/2, \\
V_1(r) &= V(r), \\
V_2(x,y) &= V(r)-F \big( \cos(\delta)x+\sin(\delta)y \big),
\label{V2}
\end{align}
where $R_w \gg 1$ and $U_w \gg 1$ are the radius and strength of the wall potential, respectively. The central potential $V(r)$ is essentially $0$ inside the trap and then it increases rapidly to the asymptotic $U_w$ around the radius $R_w$. The bright soliton is the one we actually manipulate, which experiences an additional linear potential, i.e., the second term of Eq.~(\ref{V2}). Here, $F>0$ is the force magnitude and $\delta$ is the force angle. In this work, we use $R_w=40, U_w=100$, and the force is directed towards the negative $x$-axis, i.e., $\delta=\pi$ unless specified otherwise.

It is expected from symmetry that there exist stationary VB states at the trap center when the driving force is absent, i.e., $F=0$. Assuming stationary states of the form $\psi_j(x,y,t)=\psi_j^0(x,y)\exp(-i\mu_j t)$, we get the stationary state equation:
\begin{eqnarray}
-\frac{1}{2} \nabla^2 \psi_j^0 + V \psi_j^0 + \left(\sum_{k=1}^2 g_{jk}| \psi_k^0 |^2\right) \psi_j^0 = \mu_j \psi_j^0,
\label{GPE}
\end{eqnarray}
where $\mu_1$ and $\mu_2$ are the chemical potentials of the respective components.

\begin{figure}
\includegraphics[width=\columnwidth]{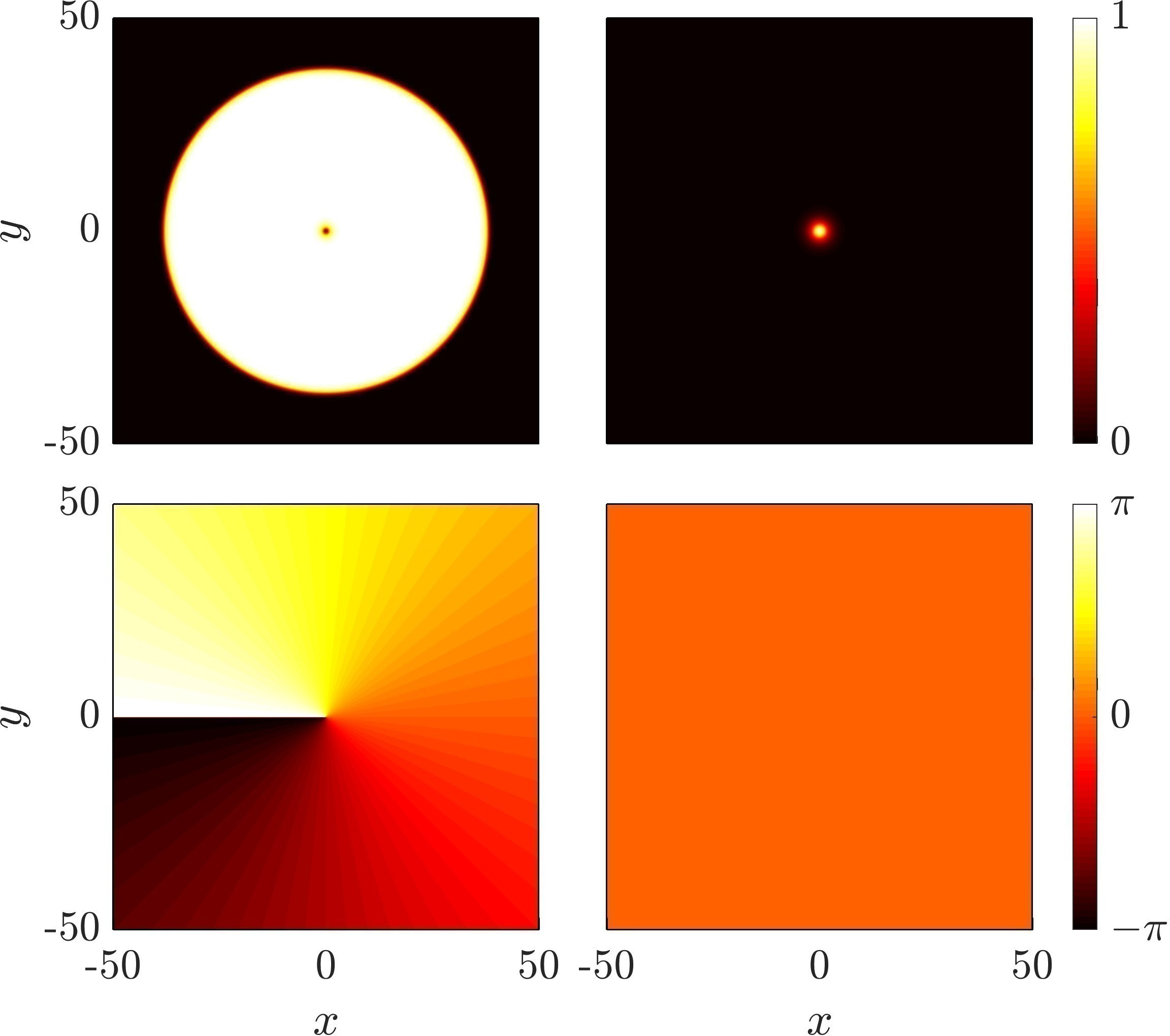}
\includegraphics[width=\columnwidth]{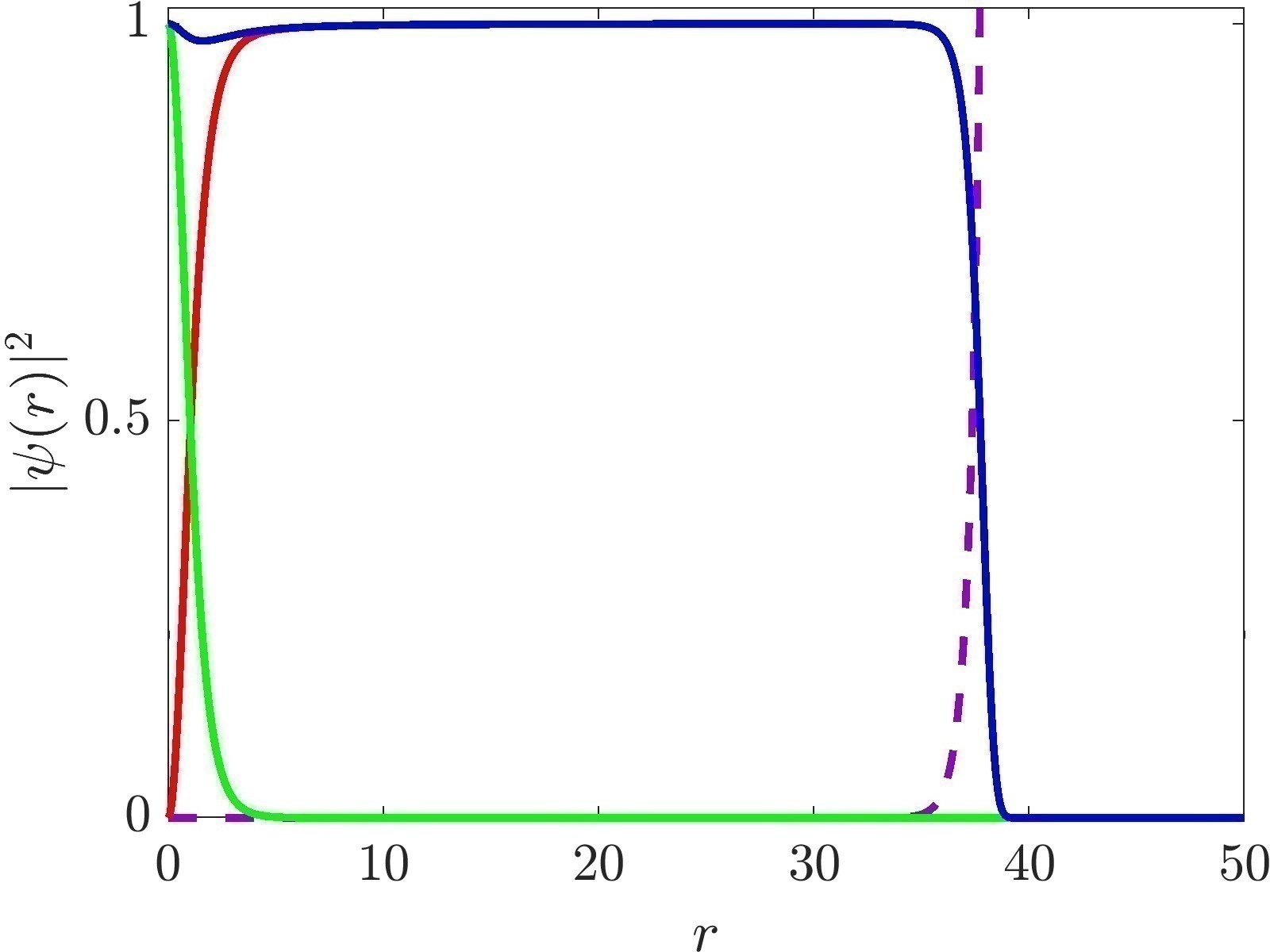}
\caption{
The top quartet depicts the stationary vortex-bright profile at $\mu_1=3, \mu_2=1.7563$, featuring approximately a constant total density in the bulk of the condensate. The top row is the magnitude $|\psi|$, the second row is the phase arg$(\psi)$, the left column is the vortex state $\psi_1$, and the right column is the bright state $\psi_2$. In the bottom panel, the radial densities of the two components $|\psi_1|^2$, $|\psi_2|^2$, as well as the total density $|\psi_1|^2+|\psi_2|^2$ are shown in red, green, and blue solid lines, respectively. The dashed line is the trapping potential, rendering the condensate finite.
}
\label{state}
\end{figure}

Our numerical setup includes identifying numerically exact stationary states using the finite element method for a spatial discretization and the Newton's method towards convergence, conducting the Bogolyubov-de Gennes stability analysis, and finally integrating the dynamics using the regular fourth-order Runge-Kutta method. Our VB soliton is numerically continued in a controlled manner from its underlying linear limit in a harmonic potential \cite{Wang:MDDD}. In this setting, the linear limit is a vortex state $(|10\rangle+i|01\rangle)/\sqrt{2}$ coupled with the ground state $|00\rangle$ with chemical potentials $2$ and $1$, respectively, where $|n_x n_y\rangle$ is a linear oscillator state in Cartesian coordinates. First, the VB is continued in chemical potentials from the linear limit to a nonlinear regime in a harmonic trap.  Next, the VB is continued from the harmonic trap to the hard-wall potential of Eq.~\eqref{Vr} using an interpolation of the potentials, keeping the chemical potentials fixed. Finally, the chemical potentials are fine tuned such that the total density is approximately uniform inside the trap in order to minimize radiations upon driving. Our optimization finally yields $\mu_1=3, \mu_2=1.7563$, and this state is depicted in Fig.~\ref{state}. This VB state is fully spectrally stable, and in fact no unstable mode is observed throughout the entire continuation, showing the robust nature of the state \cite{PK:VB}. In addition to the VB state, we have also similarly prepared a ground state where $\psi_1$, at again  $\mu_1=3$, does not contain the central vortex structure and $\psi_2=0$. When a VB configuration off the center is needed, the VB state is imprinted on top of the ground state at the desired position. In fact, multiple VBs can be imprinted as long as they are well separated and far away from the boundary. On the dynamics, a typical simulation runs up to $t=20000$, covering a few oscillation periods. Therefore, our simulations are computationally demanding due to the large domain size and the long integration time. The dynamics is very robust, and we have checked that even a simple imprinting on top of the aforementioned ground state using hyperbolic functions $(\tanh(r)\exp(i\theta)$ and $\sech(r))$ works well.

Our main observables are the VB position and the energies of different forms. Since in our dynamics the bright soliton follows very closely the vortex motion without significant deformations, we use for simplicity the center of mass of the bright soliton $(x_c, y_c)$, or $(x, y)$ when no confusion arises, to characterize the VB dynamics. In addition, we also measure the following energies:
\begin{align}
E_T &= E_K + E_P + E_I, \\
\label{EK}
E_K &= \iint \left( \frac{|\vec{\nabla} \psi_1|^2}{2} + \frac{|\vec{\nabla} \psi_2|^2}{2} \right) d^2\vec{r}, \\
E_P &= \iint \left(V_1|\psi_1|^2 + V_2 |\psi_2|^2 \right) d^2\vec{r}, \\
E_I &= \iint \left(\frac{g_{11}}{2} |\psi_1|^4 +g_{12}|\psi_1|^2|\psi_2|^2 +\frac{g_{22}}{2} |\psi_2|^4 \right) d^2\vec{r},
\end{align}
where the total energy $E_T$ is the sum of the kinetic energy $E_K$, potential energy $E_P$, and the interaction energy $E_I$ \cite{Panos:book}. Given a fixed force, the system is Hamiltonian and $E_T$ is a constant of motion. Finally, we also measure the number of atoms of each component $N_j=\iint |\psi_j|^2 d^2\vec{r}$, and they are separately conserved during the entire evolution independent of the force.


\section{Results}
\label{results}

We start by presenting the basic properties of the driving-induced VB oscillation dynamics for a series of forces on the VB initially at rest at the origin, and then we turn to the theoretical analysis. The motivation of this presentation order is that it should be helpful to build some intuitions before discussing the more abstract theory. The theory is formulated from the energy perspective, and also the use of a virtual image vortex.  From the theoretical understanding, we make further predictions to engineer the VB dynamics, which are subsequently confirmed numerically. It is worth mentioning that the dynamics and the associated engineering techniques presented herein are very flexible and robust.

\subsection{Basic properties of the VB dynamics}

\begin{figure*}
\centering
\subfigure[]{\includegraphics[width=0.28\textwidth]{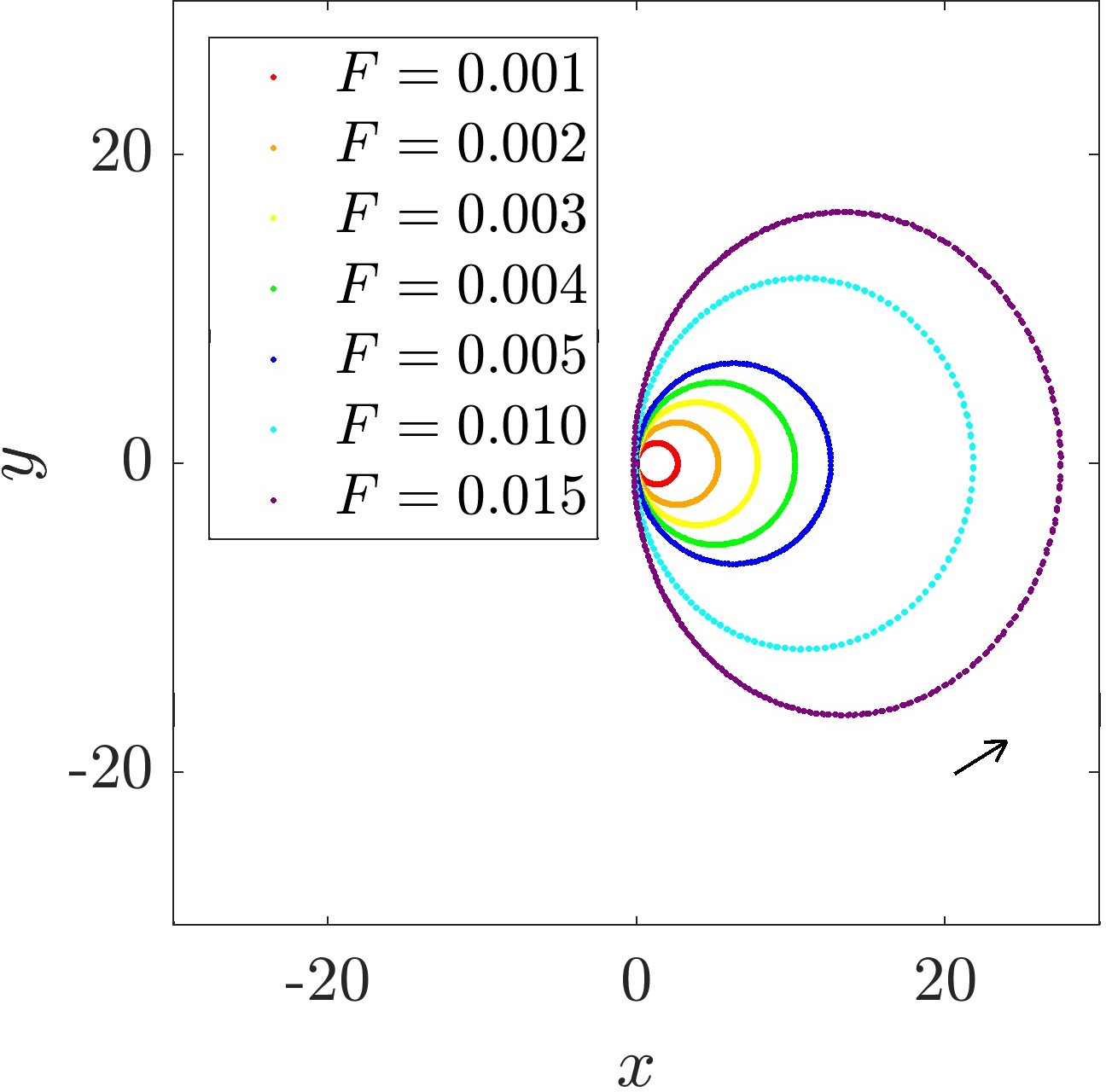}\label{trajectory}}
\subfigure[]{\includegraphics[width=0.34\textwidth]{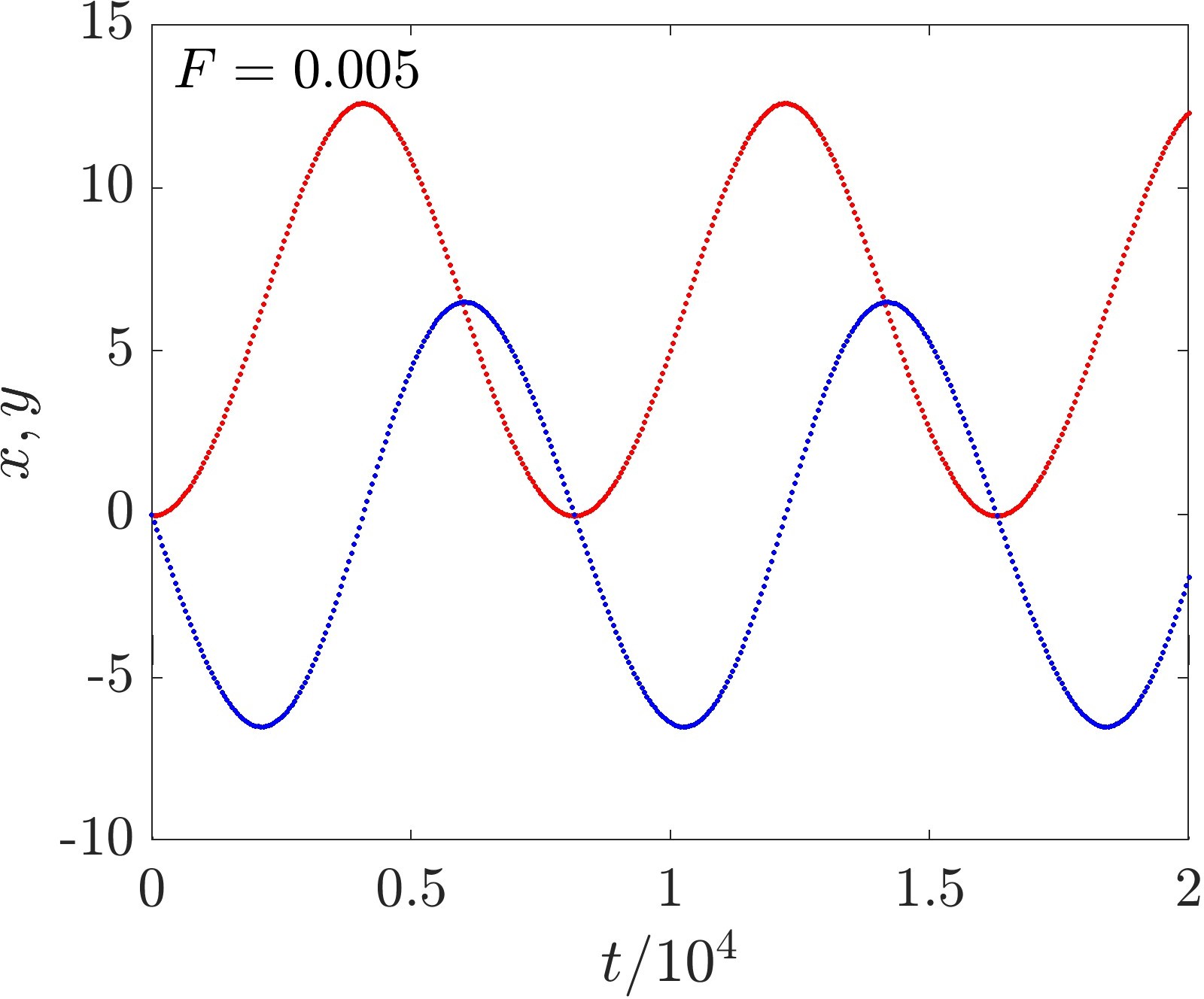}\label{ellipse}}
\subfigure[]{\includegraphics[width=0.345\textwidth]{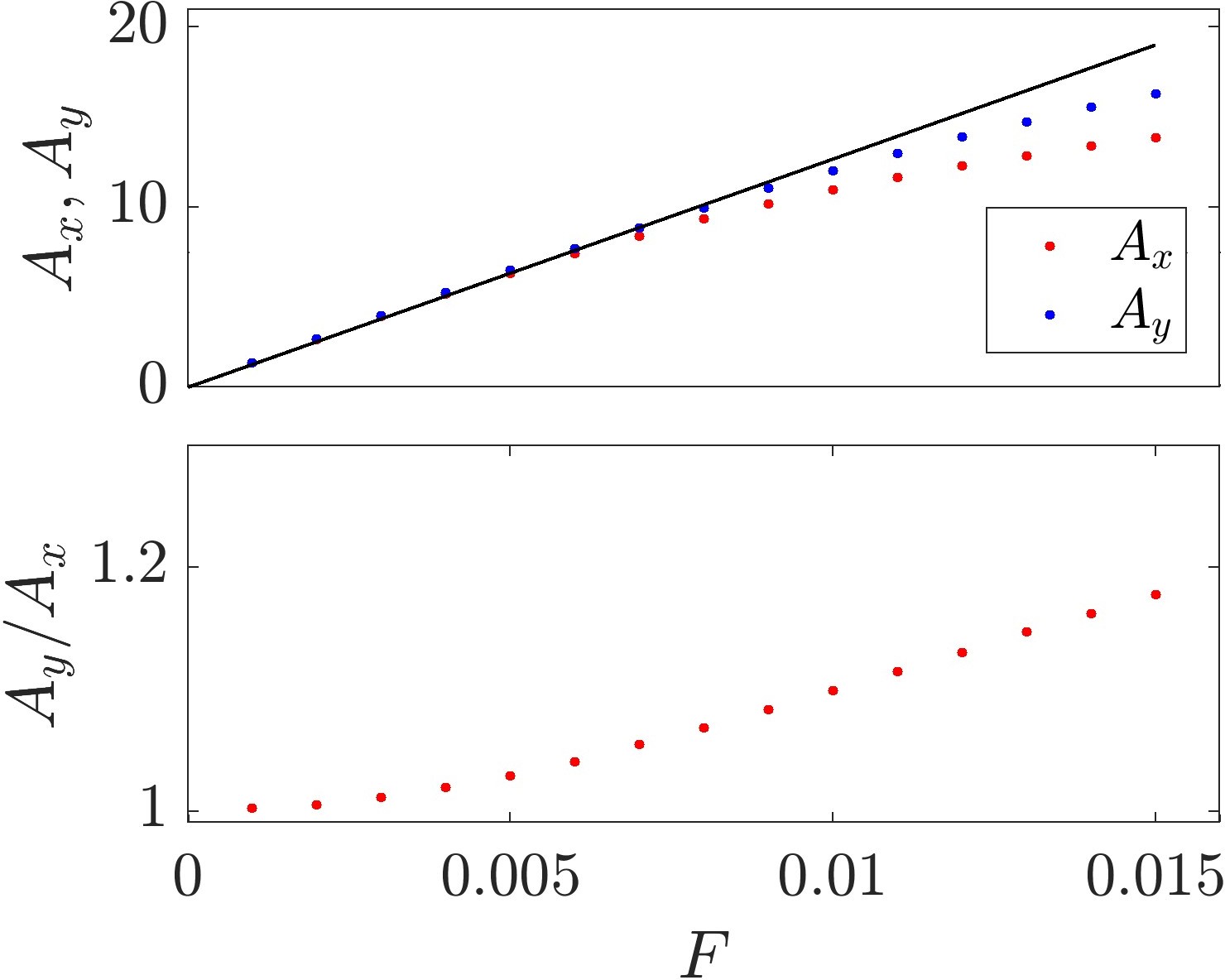}\label{amplitudes}}
\subfigure[]{\includegraphics[width=0.322\textwidth]{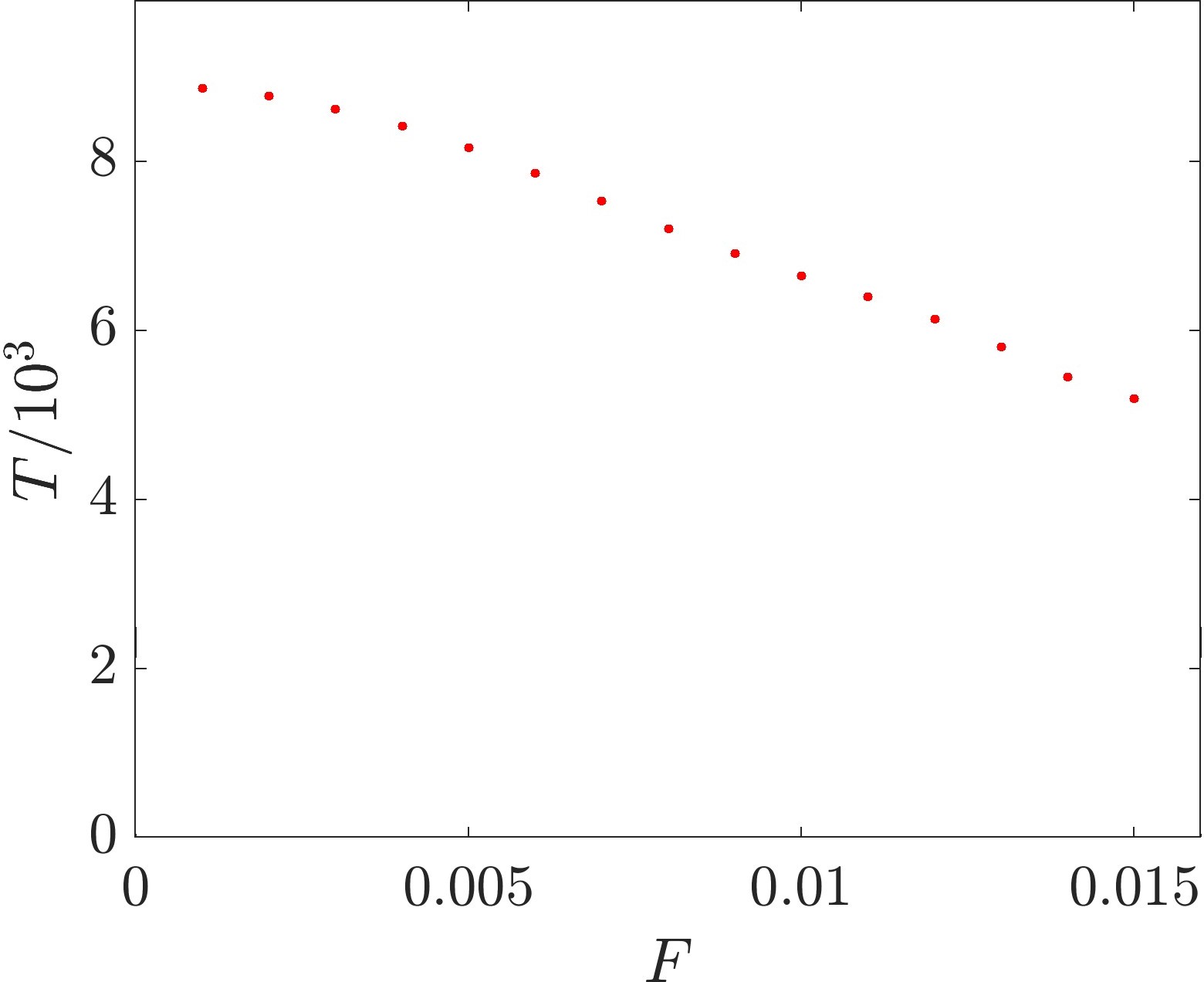}\label{periods}} 
\subfigure[]{\includegraphics[width=0.34\textwidth]{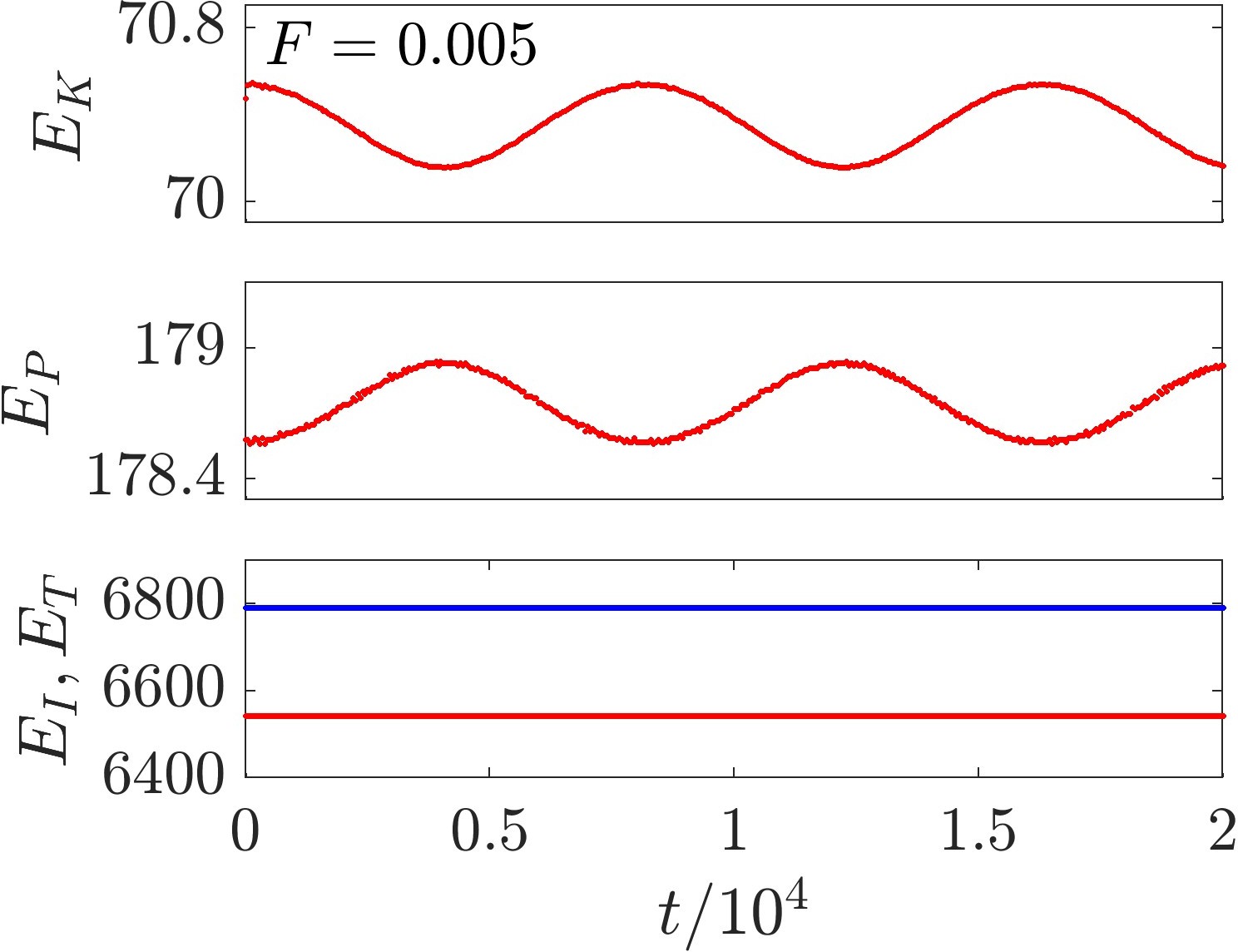}\label{energy}} 
\subfigure[]{\includegraphics[width=0.328\textwidth]{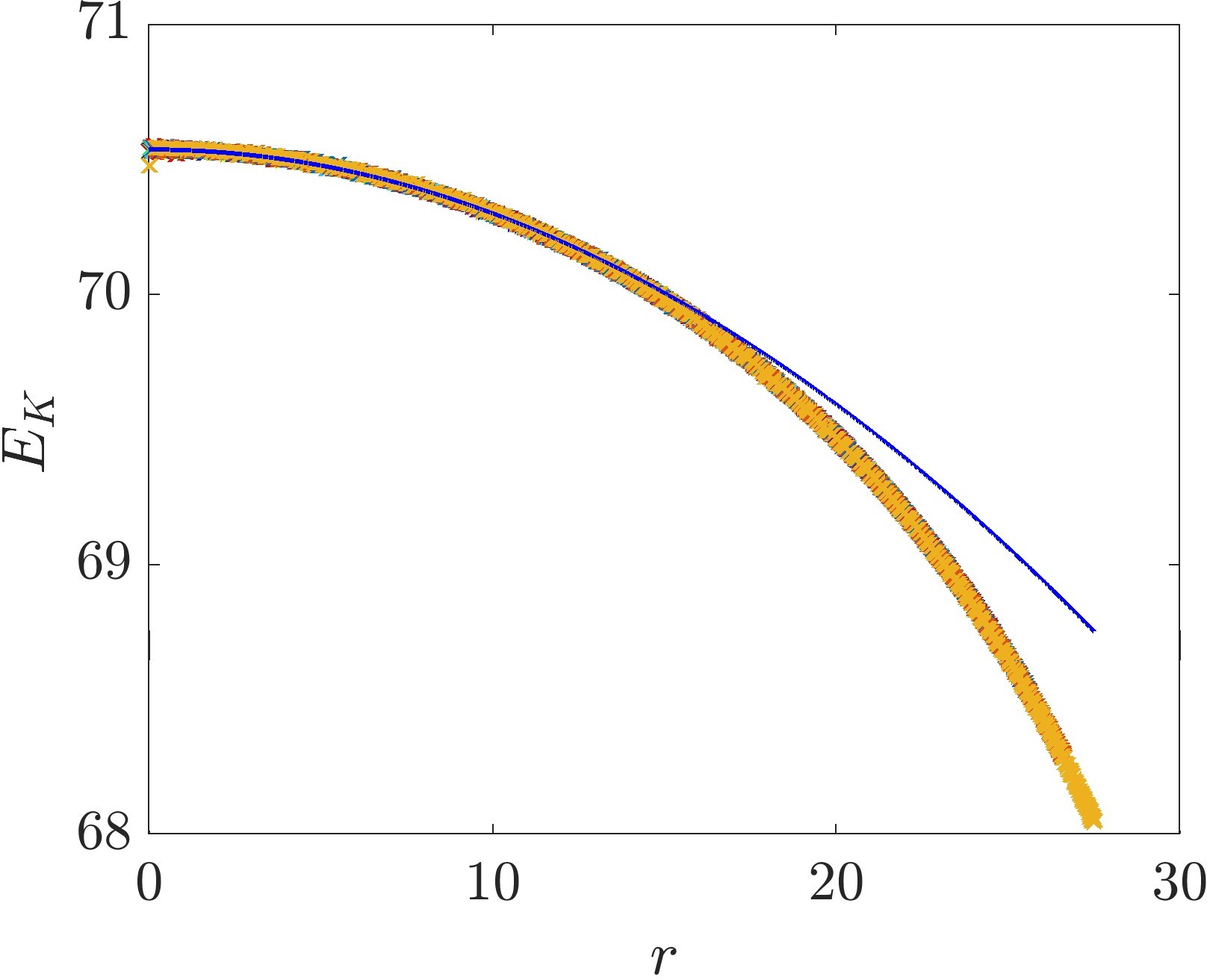}\label{disp}} 
\caption{
Panel (a): The VB initially at $(0,0)$ executes periodic oscillations upon driving on the bright component along the negative $x$-axis, the orbits are larger for larger forces. The orbits are circular in the small force limit ($F \lesssim 0.005$) and become slightly elliptical when the driving force increases. The motion is harmonic (b) and the amplitudes, linear in the small force limit, increase with increasing force, and the black line is from the theory [Eq.~\eqref{AF}] (c). The precessional period is approximately a constant, and decreases slightly with increasing $F$ (d). From the energy perspective, the system exchanges its kinetic and potential energies periodically. In our setting, the interaction energy and the total energy are constants of motion (e). The final panel (f) depicts the kinetic energy of the VB against its displacement from the center. The function is universal for different forces of (a), suggesting that the VB kinetic energy depends on its displacement due to the long-range nature of its phase winding. The blue curve is a parabolic fit using the five smallest force values, i.e., for small oscillations.
}
\label{VBdynamics}
\end{figure*}

The trajectories of the VB upon driving for a series of forces are depicted in Fig.~\ref{trajectory}. The trajectories are essentially circular for small forces $F \lesssim 0.005$, and the amplitude grows approximately linearly with $F$. Similar to the DB dynamics, the VB initially propagates \textit{against} the external potential. Different to the DB dynamics, the VB carries out a genuinely 2d uniform circular motion, while the DB inevitably executes a 1d dynamics which is an AC harmonic oscillation. In addition, the scaling properties differ significantly from the DB ones. In the DB oscillation, both the amplitude and the period are inversely proportional to the force. In the VB oscillation, the amplitude is proportional to the force instead, and we shall discuss soon that the period to leading order does not depend on the force, suggesting the two dynamics are very distinct in nature. The vortex, which is positively charged, propagates counterclockwise, and the oscillation is harmonic, as shown in Fig.~\ref{ellipse}.

An additional major difference with the DB dynamics is that the VB soliton does not deform significantly, i.e., the VB remains a point-like particle during the evolution. Since this is likely natural and as expected, we shall not present the relevant configurations here for simplicity. By contrast, the DB waveform evolves significantly during its AC oscillation because its waveform depends on its travelling velocity, e.g., a DB widens and meanwhile becomes shallower when it speeds up. On the other hand, a vortex must keep its density vanishing at the core, because of the phase singularity. This difference has an important consequence below in formulating a theory, it is necessary to examine the DB internal structure to understand its dynamics but this is not the key aspect for the VB oscillation.

Next, we quantify the VB precessional dynamics further using some helpful observables. Figure~\ref{amplitudes} shows the amplitude is proportional to the force in the small force limit. Here, both the amplitudes $A_x$ and $A_y$ along the $x$ and $y$ directions, respectively, are measured, and the trajectories become increasingly elliptical with $A_y \gtrsim A_x$ when the force is increased above $F \gtrsim 0.005$. Here, $A_x$ and $A_y$ are also the half lengths of the minor and major axes, respectively. Nevertheless, the linear scaling in the small force limit is evident. On the other hand, the period $T$ is approximately a constant of motion to leading order, as illustrated in Fig.~\ref{periods}. It is worth noting that the decrease of the period as the force or the precessional amplitude increases is a very common feature of vortex dynamics \cite{Wang:VX}.

Finally, we look at the evolution of the energies. A typical result of $F=0.005$ is shown in Fig.~\ref{energy}. The kinetic energy and the potential energy exchange periodically. The interaction energy is virtually a constant of motion. The total energy is conserved as expected. These features are similar to the DB oscillation \cite{Lichen:DB}. The potential energy oscillation is as expected from the VB dynamics, as it is due to the bright component and it is clearly a consequence of the harmonic oscillation the VB soliton. The constant interaction energy is likely because the VB does not deform significantly during the evolution. Similar features are found for all other forces. Figure~\ref{disp} shows a crucial dispersion relation on the kinetic energy of the VB as a function of its distance to the center of the condensate. Interestingly, the data from the different trajectories of Fig.~\ref{trajectory} collapse together tightly on a single universal curve. The blue curve here is a parabolic fit for the small-amplitude or the small force limit using all the data up to $F=0.005$. This correlation is a key step for appreciating the VB dynamics at an intuitive level, as we shall turn to in the next.

\subsection{Theoretical analysis}

It might be tempering to study the VB dynamics using a similar variational approach following the successful treatment of the DB dynamics. However, after much efforts of modelling and examination, this approach becomes increasingly tedious and unpromising. As mentioned earlier, it seems that the very weak \textit{internal} VB deformation is not key to its dynamics. The VB remains quite symmetric without apparent widening and shrinking throughout the evolution. It is highly important to realize that the VB is not entirely a localized soliton due to its long-range phase winding, despite that its density defect is localized. This suggests that the VB can ``feel'' the wall and its energy should be sensitive to its position in spite of the wall radius $R_w=40$ is much larger than the vortex core size which is $O(1)$. In fact, this is our motivation to examine the $E_K(r)$ relation in Fig.~\ref{disp}. By contrast, a DB virtually cannot ``feel'' the wall as long as the wall is far away \cite{Wang:DSE}. Indeed, we have verified this feature using a similar 1d wall potential, we find that the DB dynamics \cite{Lichen:DB} is completely analogous to that of the fully homogeneous system.


As the VB is a stationary state without driving at the center, the center therefore corresponds to a local energy extrema. Considering also the rotational symmetry, we expect the VB kinetic energy should take the form $E_K(r)=\alpha-\beta r^2/2$. 
Here, the phenomenological parameters $\alpha>0, \beta>0$ such that the origin is a local maximum, because as the vortex is displaced away from the center, the condensate contains less atoms with high rotating speed and more atoms with low rotating speed. Note that the superfluid speed is proportional to the gradient of the phase and therefore inversely proportional to the distance to the core of the vortex. The expectation is confirmed in the $E_K(r)$ relation in Fig.~\ref{disp}. As mentioned earlier, the blue line is a quadratic fit using data up to $F=0.005$ where the VB orbit remains close to circular. We estimate from the fit $\alpha=70.5391$ and $\beta=0.0047$. Moreover, we have checked that the kinetic energy is indeed primarily from the vortex part, the bright soliton kinetic energy, second term of Eq.~\eqref{EK}, only contributes typically less than $2\%$ to the total kinetic energy in our study.

The potential energy of the bright component is straightforward as $E_{\mathrm{PB}}=N_BFx$, where the bright atom numbers $N_B \approx 5.9543$ and it is calculated numerically from the initial state. This suggests that $E_{\mathrm{PB}}/F=N_B x$ is another universal function independent of the force with a slope of $N_B$. This is confirmed numerically and omitted here for clarity.

Finally, we put the two pieces together and derive the equation of motion from the VB energy, which is a constant of motion:
\begin{eqnarray}
E_{\mathrm{VB}} &=& E_K(r)+E_{\mathrm{PB}}(x), \\
 &=& \alpha - \beta(x^2+y^2)/2+N_B F x, \\
\dot{E}_{\mathrm{VB}} &=& -\beta x \dot{x} -\beta y \dot{y} + N_B F \dot{x}=0, \\
\dot{x} &=& - \gamma \beta y, \\
\dot{y} &=& \gamma(\beta x - N_B F),
\end{eqnarray}
where $\gamma$ is an arbitrary factor that cannot be determined from this phenomenological theory. We shall calculate it soon from the vortex kinematic dynamics using an image vortex \cite{VXspeed}. The physical meaning of $\gamma$ is, however, apparent. The sign is related to the precessional direction and its magnitude is related to the precessional frequency.

Solving the ODEs with the initial conditions $x(0)=y(0)=0$, we get:
\begin{eqnarray}
x &=& A \big(1-\cos(\omega t) \big), \\
y &=& -A \sin(\omega t), \\
\label{AF}
A &=& N_B F/\beta, \\
\omega &=& \gamma \beta.
\end{eqnarray}
There are two immediate interesting results from the theory, the orbits are circles and the radius is proportional to the force with a sloop given by $N_B/\beta=1.267\times10^3$. Using again the first five points, we find the fitting slope of the numerical data in Fig.~\ref{amplitudes} is $1.247 \times 10^3$, in good agreement with the theoretical prediction. 

Next, we compute $\gamma$ using the image vortex method. In this method, image vortices are introduced at appropriate positions such that the boundary conditions are satisfied, i.e., the superfluid velocities are parallel to the hard wall with no normal component. The dynamics of any vortex in the condensate follows from the sum of the contributions of all other vortices, including the image ones. In our setting of a circular geometry, there is a single image vortex of opposite charge at:
\begin{eqnarray}
\vec{r}_{\mathrm{IM}}= \dfrac{R_w^2}{r} \hat{r},
\end{eqnarray}
where $\vec{r}$ is the position of our real vortex \cite{VXspeed}. First, we can readily figure out the sign of $\gamma$. Our vortex has charge $1$, hence the image vortex has a charge $-1$. Consider the case $F=0$, and we perturb the vortex slightly to $(a,0)$ where $a>0$ is small. Then, the image vortex is at $\vec{r}_{\mathrm{IM}}=R_w^2/a \hat{i}$. The image vortex will therefore induce a velocity in the counterclockwise direction, as the vortex moves, the image vortex follows and keeps inducing a velocity in the $\hat{\theta}$ direction, the vortex thereby self-induces a counterclockwise precession through its image vortex. Therefore, $\gamma>0$. Since the superfluid velocity is inversely proportional to the distance to the vortex core, the induced velocity of the vortex due to its image is $1/(R_w^2/a-a)\hat{\theta}$. Consider small oscillations, i.e., $a$ is small, the speed is approximately $a/R_w^2$. Therefore, the precessional period is $T=2\pi a/(a/R_w^2)=2\pi R_w^2=1.005\times 10^4$, independent of $a$. This is in reasonably good agreement with our asymptotic numerical results of approximately $9\times 10^3$. The relatively ``large'' discrepancy is likely caused by the estimation of the size of the wall. Note that while $R_w$ indeed controls the wall size, the hyperbolic tangent profile has a width and the effective size of the wall is therefore slightly smaller than $R_w$. Indeed, $V(R_w)=U_w/2$ is already very large at $R_w$ [Eq.~\eqref{Vr}]. If we estimate an effective size using the value where the total density drops to half the background density, i.e., $0.5$, we obtain an effective size $R_{w\mathrm{eff}}=37.72$ and the period is then estimated as $8.938 \times 10^3$, in good agreement with the numerical results. 
An acute reader may also wonder whether the induced speed $v=1/d$, where $d$ is the distance between two vortices, needs a prefactor as, e.g., $v=B/d$. Indeed, we have conducted an additional set of simulations where two VBs of the same charge corotate around each other. Our results show that $v=1/d$ is a good approximation in our setting when the vortices are reasonably far apart $d \gtrsim 10$. This is clearly satisfied as the image vortex is far away for small oscillations.
Finally, $\gamma \approx 1/(\beta R_{w\mathrm{eff}}^2) =0.1496$.

\begin{figure}
\includegraphics[width=0.495\columnwidth]{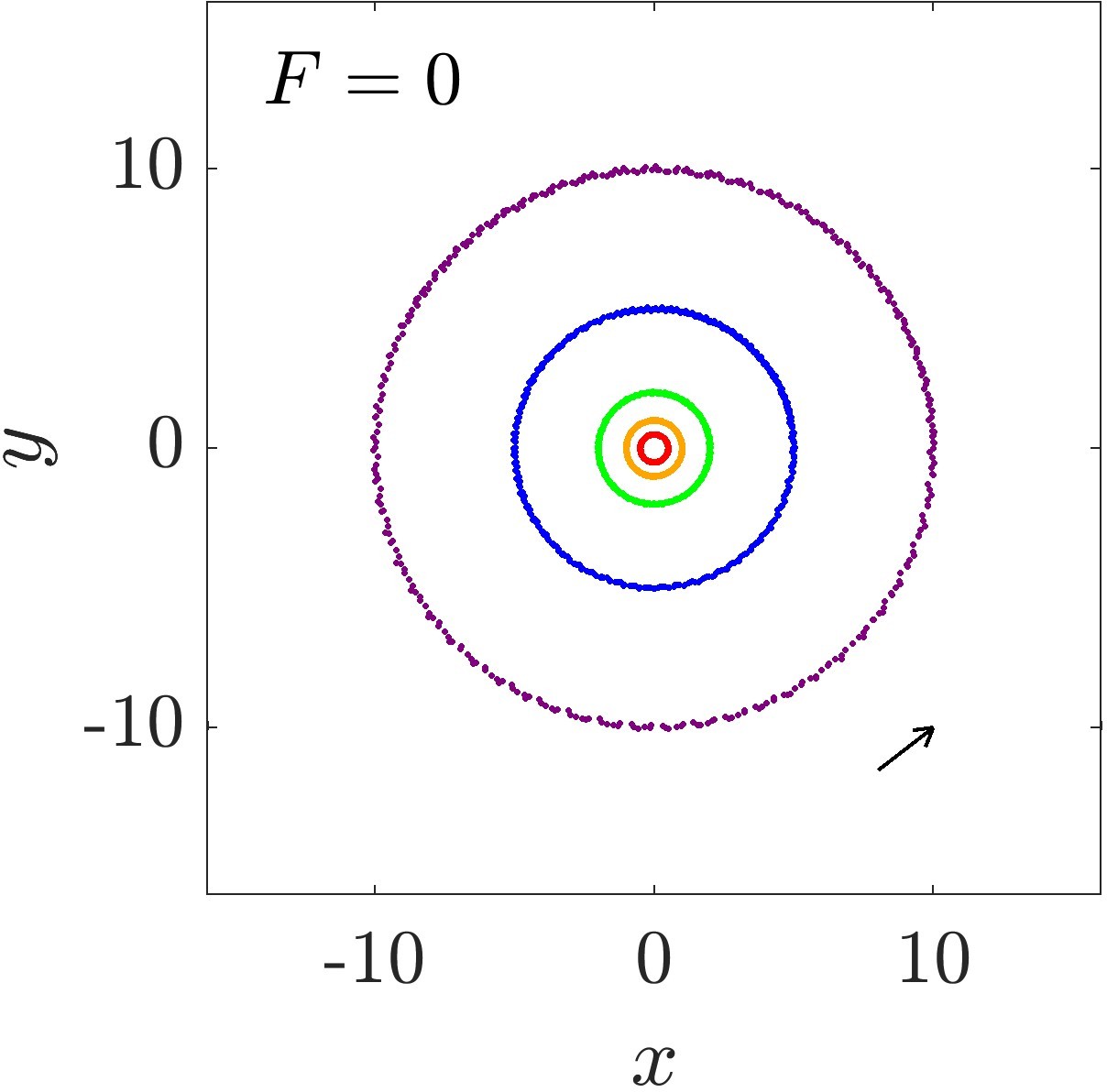}
\includegraphics[width=0.495\columnwidth]{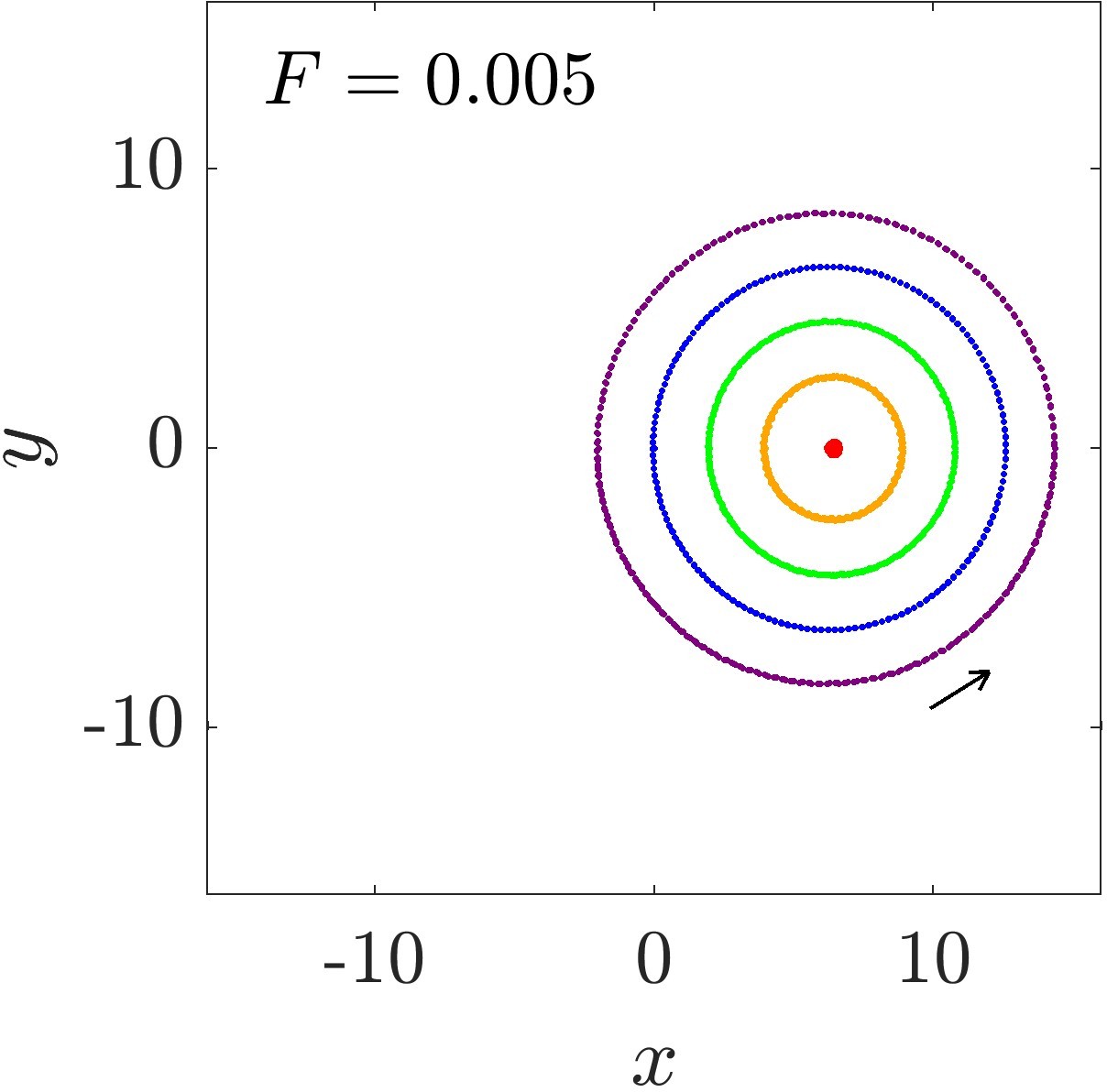}
\caption{
Two types of dynamics predicted by the theory. The left panel shows that the VB precesses when displaced from the center even if there is no force applied and the size of the condensate is pretty large. This confirms the importance of the long-range nature of the vortex phase winding. In presence of a force, the fixed point and the associated trajectories are shifted as illustrated in the right panel. The precessional directions are the same with also a similar period, in agreement with the theory.
}
\label{Shiftdynamics}
\end{figure}


The theory predicts two interesting results. When a small force is applied, there is a new fixed point at $(N_BF/\beta, 0)$, the force on the bright soliton has an effect of shifting the fixed point. Second, in the case $F=0$ as already discussed above, the VB should precess if it is shifted even if no force is applied. This is similar to a VB precession in a harmonic trap, however, our condensate is essentially homogeneous around the center. Both types of dynamics are numerically confirmed and they are illustrated in Fig.~\ref{Shiftdynamics}. The VB executes precessional motion with no force applied around the center. The new fixed point, which appears exactly at the predicted position, is particularly interesting. There are also precessional orbits around this fixed point, therefore, the trajectory of $F=0.005$ shown in Fig.~\ref{trajectory} is nothing but a particularly precessional orbit of this fixed point passing through the origin $(0,0)$. The center of each trajectory therein is a fixed point for the corresponding driving force.

\subsection{Engineering VB dynamics}

The availability of a continuously tunable fixed point and their associated precessional orbits clearly provides a highly flexible approach of engineering VB dynamics. Below, we discuss a few proof-of-principle elementary processes. By combining these elementary processes, the VB can be guided, pinned, and released at will.

First, we generalize the ODEs and the solutions to the general form:
\begin{eqnarray}
\dot{x} &=& - \gamma (\beta y+N_B F_y), \\
\dot{y} &=& \gamma(\beta x + N_B F_x), \\
\vec{r}^{\star} &=& - \dfrac{N_B}{\beta} \vec{F},
\end{eqnarray}
where $\vec{r}^{\star}$ is the fixed point of the force $\vec{F}$. In this setting, if the VB is at a position $\vec{r}_0$ at $t_0$, then the subsequent precessional evolution by solving the ODEs is given by:
\begin{eqnarray}
z = z^{\star} + (z_0-z^{\star}) e^{i\omega(t-t_0)},
\label{solution}
\end{eqnarray}
if the VB remains reasonably far away from the boundary of the wall. Here, we have used for convenience a complex coordinate $z=x+iy$ for a VB at $(x, y)$. It is easier to rewrite the ODEs to a complex form and then solve the ODE with a single but complex variable. Mathematically, the solution Eq.~\eqref{solution} is a sum of a particular solution of the inhomogeneous part (the fixed point) and a general solution of the homogeneous part, representing altogether shifted precessional dynamics around the fixed point.

\begin{figure*}
\centering
\subfigure[\ Send-Orbit-Receive dynamics]{\includegraphics[width=0.33\textwidth]{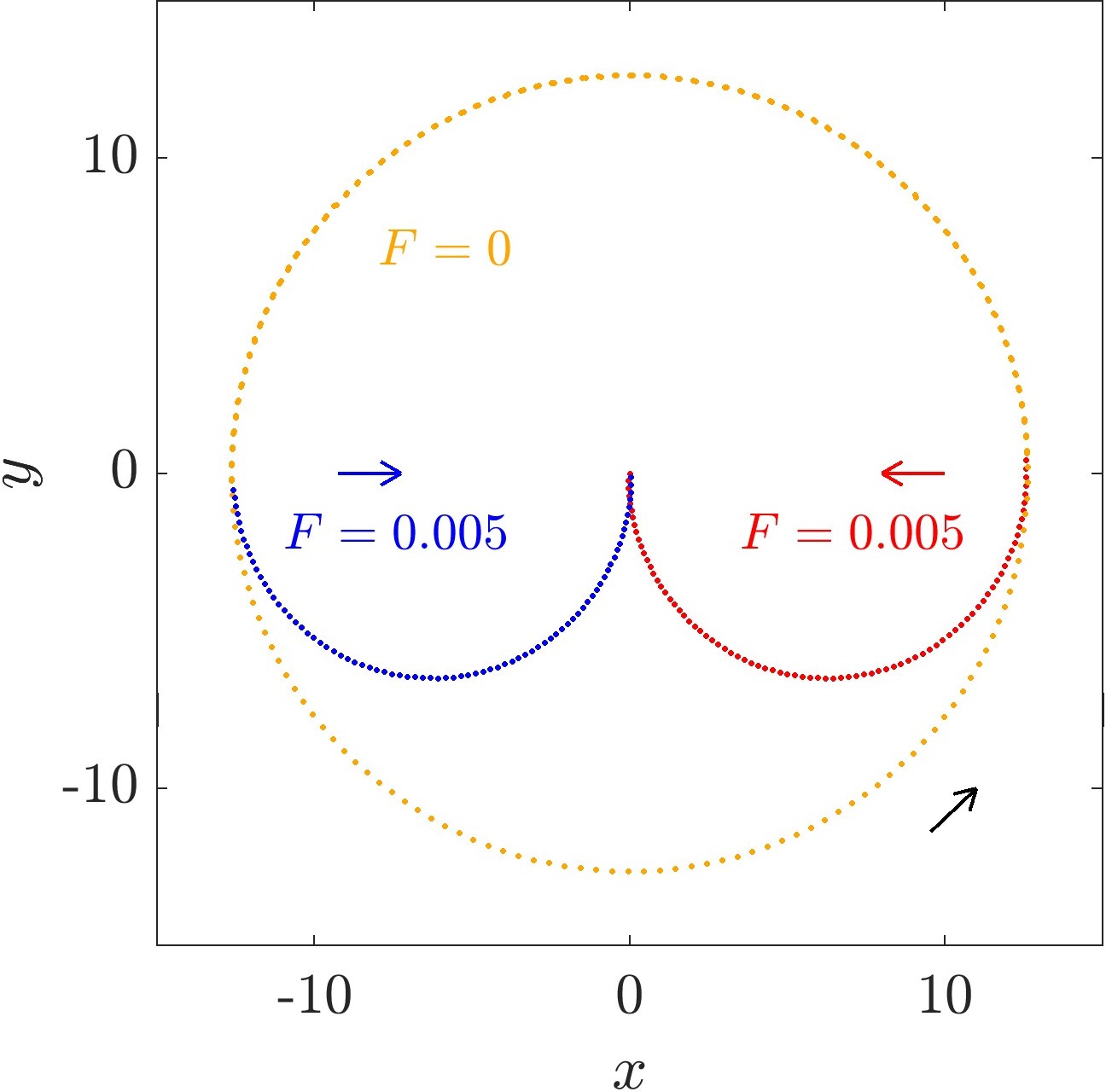}\label{Dy1}}
\subfigure[\ Pin-Orbit-Pin dynamics]{\includegraphics[width=0.33\textwidth]{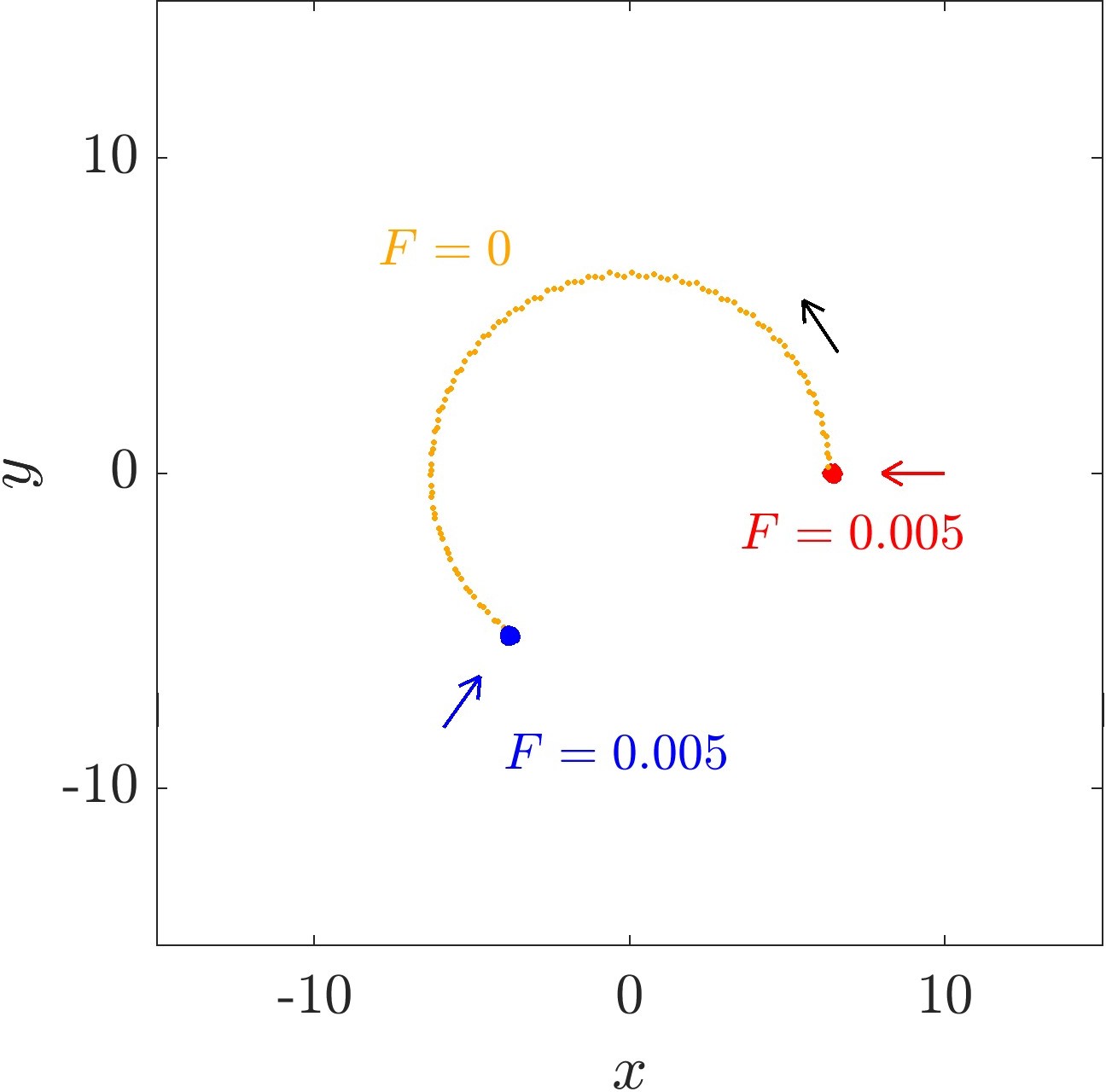}\label{Dy2}}
\subfigure[\ $A$-to-$B$ dynamics]{\includegraphics[width=0.33\textwidth]{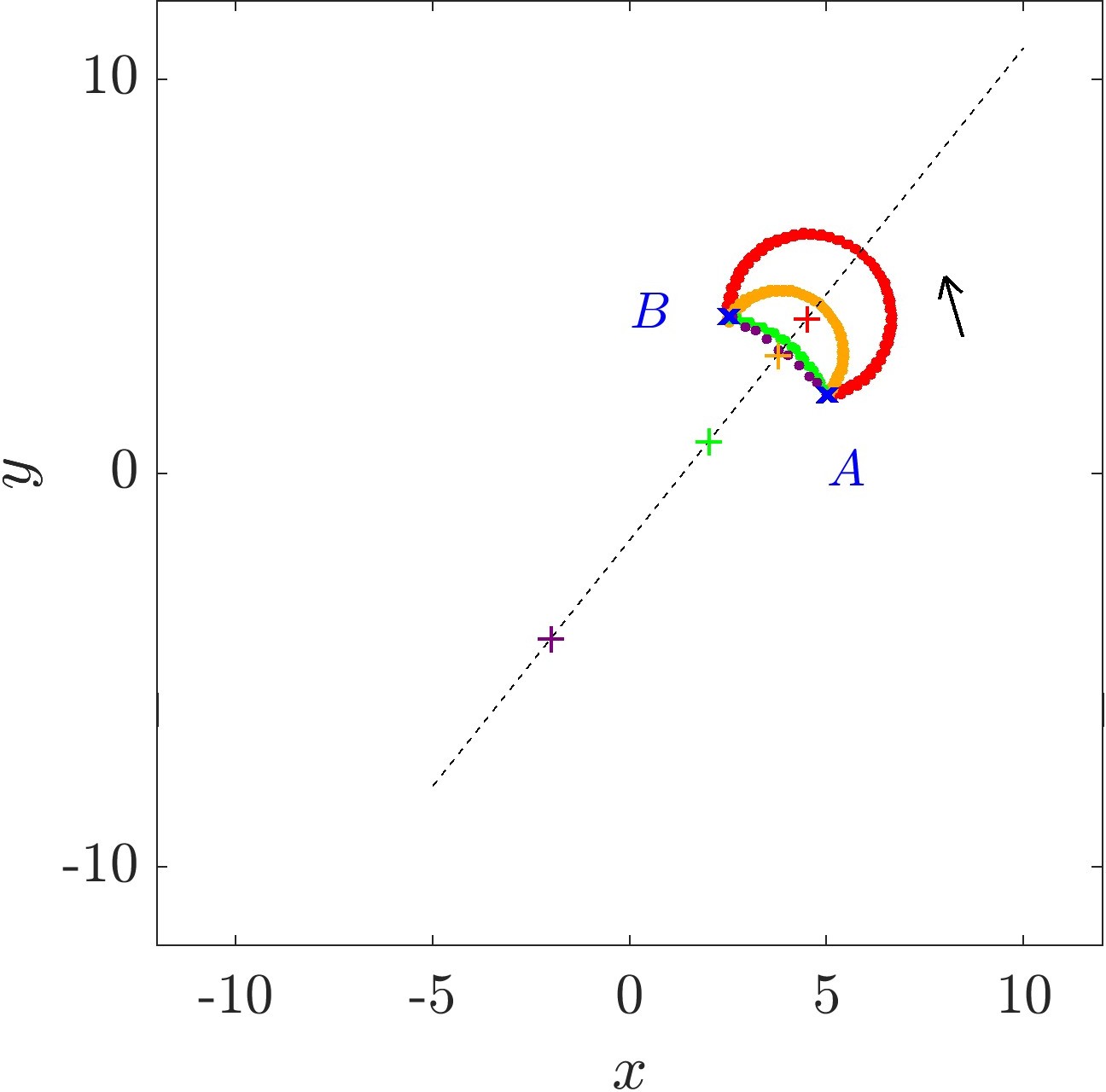}\label{Dy3}} 
\caption{
In the Send-Orbit-Receive dynamics, the VB starts from the center and is elevated to a precessional orbit under the force in red color. The force is then turned off and the VB precesses for one and a half periods until the blue force is switched on, after which the VB travels back to the origin. In the Pin-Orbit-Pin dynamics, the VB is initially pinned by the red force. It is then released and precesses until it is pinned again by the blue force. In the $A$-to-$B$ dynamics, we aim to transfer the VB from $A (5, 2)$ to a nearby $B (2.5, 4)$ and there are many ways to accomplish this transfer. The fixed point can be chosen anywhere along the perpendicular bisector, shown as the thin dashed line, provided that the trajectory stays away from the boundary. Here, four typical fixed points (the plus markers) and their trajectories (the arcs) are shown. Note that these paths trace very different angles with respect to their respective fixed points, and therefore they take very different times to complete the transfer. Please see the text for more details.
}
\label{Dynamics3}
\end{figure*}

We have designed from the above solution a total of three types of dynamical processes to illustrate the level of control on the VB dynamics:
\begin{enumerate}
\item Send-Orbit-Receive. In this dynamics, we start from a stationary VB at the center. We drive it to a desired central orbit using a force. The VB is then released and it executes precessional motion. Finally, the VB is driven back to the origin.
\item Pin-Orbit-Pin. In this dynamics, we start from a pinned VB, it is then released and continues its precession, and at a later time it is pinned again elsewhere.
\item $A$-to-$B$. This is a very elementary process. In this dynamics, we choose two arbitrary nearby points $A$ and $B$, we show that a series of orbits are available to transport the VB from $A$ to $B$. In addition, different orbits have different transport speeds.
\end{enumerate}

These dynamics are summarized in Fig.~\ref{Dynamics3} and we now discuss them in turn.
In the Send-Orbit-Receive dynamics, the VB is initially at the origin. It is then elevated to a higher orbit by the red force of magnitude $0.005$. After half a cycle at about $t=4100$, the VB reaches its maximum distance from the center $2A_x=12.6435$, and it is then released. The VB precesses afterwards without the driving force. After approximately one and a half periods $\Delta t=11850$, the VB is driven back to the origin under the blue force.

In the Pin-Orbit-Pin dynamics, the VB is initially pinned by the force (in red color) at its fixed point. It is then released at a random instant $t=16750$, after which the VB executes the precessional motion. After $\Delta t=5600$ the VB arrived at the blue point, and a pinning force (in blue color) is applied and the VB is again pinned. Here, the blue point is also randomly selected, and the VB remains pinned for a period of $\Delta t = 20000$ to the horizon of our simulation.

The $A$-to-$B$ dynamics is clearly the most generic and flexible engineering process, as if we can guide the VB from any point to any nearby point, it can follow an arbitrary trajectory. In addition, the time it takes for a small segment can also be engineered, depending on the angle the path traces with respect to its fixed point, i.e., there are multiple ways of very different time scales to transfer a VB from $A$ to $B$. Here, we examine a specific example and transfer the VB from $(5, 2)$ to $(2.5, 4)$. These two points are again randomly selected, and their separation is in fact slightly exaggerated to illustrate the idea clearly. Nevertheless, the separation remains ``small'' compared with the size of the condensate. To accomplish the transfer, we only need a precessional orbit that passes through the points $A$ and $B$. It is straightforward to see that this can be realized if the fixed point falls on the perpendicular bisector of the chord $AB$ and the arc $AB$ is the trajectory in the counterclockwise direction, provided that the trajectory does not interfere with the boundary. Note that the choice of the fixed point can be continuously tuned. If the path opens an angle $\Lambda \in (0, 2\pi)$ with respect to the fixed point, the transfer time is approximately $\Lambda T /(2\pi)$. Note that if $\Lambda$ is small, i.e., the trajectory is only a small fraction of the entire precessional orbit, the transfer time can be short despite that $T$ can be large. Four typical paths are illustrated in Fig.~\ref{Dy3}, and the coordinates of the fixed points from left to right are $(-2, -4.1875)$, $(2, 0.8125)$, $(3.75, 3)$, and $(4.5, 3.9375)$, respectively. The angles $\Lambda = 0.3444$, $1.0383$, $3.1416$, and $4.4286$, respectively. Note that the first trajectory is both more smooth for a piecewise interpolation and more efficient in time, e.g., it is about $10$ times faster than the latter two ones. This suggests that it is helpful to choose a small $\Lambda$ path when it is appropriate.

\begin{table}
\caption{
Comparison of the driving-induced AC oscillations of the DB \cite{Lichen:DB} and VB in a finite-homogeneous background. A common feature is that $E_K$ and $E_P$ exchange periodically, and both solitons initially propagate against the external linear potentials at the cost of their respective kinetic energies.
\label{table}
}
\begin{tabular*}{\columnwidth}{@{\extracolsep{\fill}} l c r}
\hline
\hline
Property  &DB &VB \\
\hline
Dimension  &1d &2d \\
Dynamics  &Harmonic oscillation &Uniform circular motion \\
$E_K$ source  &Structural deformation &Vortex shift \\
Boundary effect  &Unimportant &Important, it affects $E_K$ \\
Particle nature  &Short range &Long range \\
Amplitude &$A \propto 1/F$ &$A \propto F$ \\
Period &$T \propto 1/F$ & $T=$ const. \\
Fixed point  &Destroyed &Shifted: $\vec{r}^{\star} \propto - \vec{F}$ \\
Engineering  &Hard &Flexible and robust \\
\hline
\hline
\end{tabular*}
\end{table}

Finally, we compare the VB oscillation and the DB oscillation under a constant driving force on the bright component. A common feature is that they both initially counterpropagate against the driving potentials, and the $E_K$ and $K_P$ exchange periodically. However, there are also dramatic differences between the two dynamics. This can be traced back to the short-range nature of the DB soliton but the VB soliton has a long-range character due to its phase winding. In addition, while the counterpropagation is powered by the kinetic energy, their mechanisms are very different. The DB lowers the kinetic energy by a structural deformation, but the VB accomplishes this by off-center shifting. These differences lead to very different scaling properties, summarized in Table~\ref{table}. Finally, because the VB fixed point is shifted rather than being destroyed, as well as its robust nature during the evolution, there exists flexible and robust methods of engineering its dynamics in BECs.

\section{Conclusions and Future Challenges}
\label{conclusion}
We studied the VB precessional dynamics under a constant driving force on the bright component, finding that the fixed point and the associated precessional orbits can be shifted controllably by the force. The dynamics is explained theoretically, and we identify that the long-range nature of the VB through its phase winding is responsible for the observed dynamics. The AC dynamics provides an excellent approach of engineering the VB dynamics, as demonstrated using three prototypical examples. Finally, the dynamics is compared with that of the DB counterpart.

The driving-induced dynamics can be extended in multiple directions in the next. First, the work should be studied in the Manakov setting. From the robust nature of the VB soliton and also the recent extension of the DB AC oscillation to Manakov systems \cite{Wang:MDDD}, it is expected that the VB AC dynamics presented herein should well exist at the more experimentally relevant Manakov interactions. Indeed, our preliminary results show that the dynamics can be readily realized to the Manakov interactions, and the constant interaction energy feature appears to be quite generic in a homogeneous trap without requiring the total density is approximately uniform.
Second, multiple vortex-bright structures like the vortex-dipole-bright \cite{PK:VB} and three-dimensional filled vortical filaments like the vortex-line-bright and vortex-ring-bright configurations \cite{Wang:SO2} are particularly interesting, e.g., these structures are highly relevant in superfluid dissipation. Here, the dynamics can be considerably more versatile for at least two reasons. First, there are more available structures, e.g., the vortex-dipole-bright state comes in two types, the two bright solitons can be either in phase or out of phase \cite{PK:VB}. In addition, many more complex vortex-bright clusters are available in the literature \cite{Panos:DC2}. Second, these states are no longer point-like structures, therefore, there are different possible relative force orientations. For example, the force on the vortex-ring-bright can be applied either along the ring axis or perpendicular to it or towards a more generic direction, leading to presumably different evolution scenarios.

The DB AC dynamics should be revisited in a harmonic trap, it appears that a new fixed point may emerge due to the inhomogeneity from the energy perspective. If so, DB engineering should also be possible, despite that it may not be as flexible as that of the VB as the DB waveform depends on its velocity. Nevertheless, if a new fixed point can be tuned in a controlled manor, it is still of considerable interest in manipulating dark solitons. Dark-bright configurations are also available in higher dimensions such as the planar dark-bright, shell dark-bright, and ring dark-bright states \cite{Wang:DBS,Wang:RDS}.

Finally, the AC dynamics above has been discussed only in the context of a single constant driving force, other force fields are also interesting. For example, the force on each bright soliton can be tuned separately if they are reasonably apart. In this setting, intrinsically dynamic states like two corotating VB solitons may be controllably stopped using suitable driving forces. It is also interesting to design driving potentials for guiding a soliton following an arbitrary path and also investigate other trap geometries like asymmetric ones. Research work along some of these directions are currently in progress and will be reported in future publications.

\begin{acknowledgments}
We thank Li-Chen Zhao for helpful discussions. We gratefully acknowledge supports from the National Science Foundation of China under Grant No. 12004268, the Fundamental Research Funds for the Central Universities, China, and the Science Speciality Program of Sichuan University under Grant No. 2020SCUNL210.
We thank the Emei cluster at Sichuan university for providing HPC resources.
\end{acknowledgments}

\bibliography{Refs}

\end{document}